\documentstyle{aa}     
\def\ut#1{\mathop{\vtop{\ialign{##\crcr
     $\hfil\displaystyle{#1}\hfil$\crcr\noalign
     {\kern1pt\nointerlineskip}\hbox{$\hfil\sim\hfil$}\crcr
     \noalign{\kern1pt}}}}}

\def\undersymbol#1#2{\mathop{\vtop{\ialign{##\crcr
     $\hfil\displaystyle{#2}\hfil$\crcr\noalign
     {\kern1pt\nointerlineskip}\hbox{$\hfil#1\hfil$}\crcr
     \noalign{\kern1pt}}}}}

\def\arcmin{^{\prime}}
\def\degr{^0}
\begin{document}
\title{
   Kerr black holes as retro-MACHOs}
\author{
    F. De Paolis
\inst{1},
       A. Geralico
\inst{1,2},
 G. Ingrosso
 \inst{1},
 A. A. Nucita \inst{1} \and
 A. Qadir\inst{3,4}} \offprints{F. De Paolis} \institute{
Dipartimento di Fisica, Universit\`a di Lecce, and {\it INFN},
Sezione di Lecce, Via Arnesano, CP 193, I-73100 Lecce, Italy \and
International Center for Relativistic Astrophysics - I.C.R.A.,
University of Rome ``La Sapienza'', I-00185 Rome, Italy\and
Department of Mathematics, Quaid-i-Azam University, Islamabad,
Pakistan \and Department of Mathematical Sciences, King Fahd
university of Petroleum and Minerals, Dhahran 31261, Saudi Arabia}
\date{Received 14 May 2003/ Accepted 17 October 2003}
\authorrunning{De Paolis et al.}
\titlerunning{Kerr black holes as retro-MACHOs}

\abstract{Gravitational lensing is a well known phenomenon
predicted by the General Theory of Relativity. It is now a
well-developed observational technique in astronomy and is
considered to be a fundamental tool for acquiring information
about the nature and distribution of dark matter. In particular,
gravitational lensing experiments may be used to search for black
holes. It has been proposed that a Schwarzschild black hole may
act as a retro-lens (Holz \& Wheeler \cite{hw}) which, if
illuminated by a powerful light source (e.g. the Sun), deflects
light ray paths to large bending angles so that the light may
reach the observer.  Here, by considering the strong field limit
in the deflection angle and confining our analysis to the black
hole equatorial plane, we extend the Holz-Wheeler results to
slowly spinning Kerr black holes. By considering the Holz-Wheeler
geometrical configuration for the lens, source and observer we
find that the inclusion of rotation does not substantially change
the brightness of the retro-lensing images with respect to the
Schwarzschild case. We also discuss the possibility that the next
generation space-based telescopes may detect such retro-images and
eventually put limits on the rotational parameter of the black
hole.

} \maketitle

\section{Introduction}

Gravitational lensing of electromagnetic waves is a
well-understood phenomenon (see e.g. Schneider et al. \cite{sef}
for a comprehensive treatment) predicted by the theory of General
Relativity. In the last years, gravitational lensing has been also
used as an observational tool to investigate the mass distribution
both in galaxy clusters (using distant QSOs as sources, Canizares
\cite{canizares1982}) and in galactic halos. In the latter case,
the phenomenon, which is called gravitational microlensing,  led
to the discovery of Massive Astrophysical Compact Halo Objects
(MACHOs) in the halo of our Galaxy (Alcock et al.
\cite{alcock1993}, Aubourg et al. \cite{aubourg1993}).

As a dark object moves across the source-observer line of sight it
acts as a gravitational lens leading to the formation of two
unresolved images of the source and to an overall source light
magnification. Despite the low probability of observing such
events, Paczy\'{n}sky (\cite{pacz}) showed that the continuous
observation of at least $\simeq 10^6$ stars implies a
non-negligible chance of detecting a microlensing event towards
the LMC.

Since the early 1990s several collaborations (MACHO (see e.g.
Alcock et al. \cite{alcock1993}), EROS (Aubourg et al.
\cite{aubourg1993}) and OGLE (Stanek et al.  \cite{ogle})) have
been monitoring millions of stars towards the LMC, the SMC, as
well as towards the galactic center detecting hundreds of
microlensing events due to MACHOs along the line of sight.
Unfortunately, the physical nature of MACHOs is still unknown,
since the observed light curves of a microlensing event can be
reproduced by models depending on a number of free parameters
(mass, distance and transverse velocity of the lens). So,
different populations of objects such as  brown dwarfs, white
dwarfs, main sequence stars and black holes may be consistent with
microlensing observations. In particular, it is generally accepted
that a population of black holes may exist in the Galaxy and it
has been claimed that these compact objects have already been
detected in microlensing surveys (Quinn et al. \cite{quinn},
Bennett et al. \cite{bennett}, Agol et al. \cite{agol}). At least
6 extremely long events detected by the MACHO, GMAN and MPS
collaborations towards the galactic bulge (Bennett et al.
\cite{bennett}) exhibit very strong microlensing parallax signals
which lead to mass estimates up to $\simeq 10 ~M_{\odot}$. Since
the estimated upper limits on the absolute lens brightness are
always less than $\sim 1~L_{\odot}$, this favors that the stellar
black hole hypothesis (Mao et al. \cite{mao}) and suggests that a
substantial fraction of the galactic lenses may be massive stellar
remnants.

Despite some experimental uncertainties, the theory of
gravitational lensing, which was originally developed in the weak
field approximation, has successfully explained all gravitational
lensing observations from the giant blue luminous arcs in the rich
cluster of galaxies (Lynds \& Petrosian \cite{lp}, Soucail et al.
\cite{soucail}) to the first Einstein ring MG1131+0456 at redshift
$z\simeq 1.13$ (Hewitt et al. \cite{hewitt}) to microlensing
events detected towards the SMC, LMC, the galactic bulge and the
M31 galaxy (see e.g. Alcock et al. \cite{alcock2000}).

Moreover, by using the next generation of high resolution imaging
telescopes, it would be possible to test the theory of
gravitational lensing in the presence of strong gravitational
fields since the inspection of the images formed will allow us to
investigate regions very close to the surface of compact massive
objects such as neutron stars and black holes.

In a very interesting paper Holz \& Wheeler (\cite{hw}) proposed
that a Schwarzschild black hole may act as retro-lens or
retro-MACHO. This means that if the black hole is illuminated by a
powerful light source, photons, due to its strong gravitational
field, can be deflected by large bending angles. In the case of
perfect alignment of source, black hole and Earth (with the Earth
in the middle), the bending angles are odd multiples of $\pi$
leading to a series of circular rings.

However, black holes would be characterized by a non-zero
intrinsic angular momentum, which breaks the Schwarzschild
spherical symmetry and affects the gravitational field around the
compact object.  Thus, we expect a modification in the
phenomenology of the gravitational lensing. Here, we investigate
the retro-lensing phenomenon for slowly rotating Kerr black holes
considering photon trajectories close to the black hole equatorial
plane. Our treatment follows the second order expansion  (given by
Bozza \cite{bozza2}) of the large deflection angles of light rays
winding around the retro-MACHO in the impact parameter $b$.

This paper is structured as follows. In Sect. 2 we briefly review
the Schwarzschild retro-MACHO lensing results obtained by Holz \&
Wheeler (\cite{hw}). In Sect. 3 we extend the Holz \& Wheeler
(\cite{hw}) results to Kerr black holes . In Sect. 4 we
investigate different geometries among the lens, the source and
the observer and discuss the possibility that the next generation
high resolution telescopes may have the capabilities necessary to
detect such lensing events.

\section{Schwarzschild retro-MACHOs}
The theory of General Relativity predicts that the gravitational
field of a massive object deflects light rays of a background
source, if the impact parameter is sufficiently close to the
massive object. This phenomenon is referred to as gravitational
lensing and the massive object causing a detectable deflection is
called a gravitational lens.

In gravitational microlensing, photons from a distant source
suffer a very small angular deflection and the weak field
approximation always works satisfactorily. However, strong field
effects may be extremely important if the lens is a collapsed
compact object. In fact, very close to a neutron star or a black
hole, GR plays an important role and the study of the relativistic
images which form (as a consequence of light deflection) allows us
to investigate the regions close to the event horizon. The light
bending angle in the presence of a static black hole is not
limited to small angles but may reach large values ($\pi$ and odd
multiples of $\pi$ if source, observer and retro-MACHO are
perfectly aligned, see Holz \& Wheeler \cite{hw}) so that photons
emerge in the direction of the source itself. Therefore, a
Schwarzschild black hole, if illuminated by a powerful light
source, shines back with a series of micro-arcsecond rings.

In the case of perfect alignment between the source, the observer
and the lens the image of the source, in the plane orthogonal to
the line of sight and containing the source, is an annulus with
outer and inner radii corresponding to the appropriate impact
parameters for photons coming from the top and bottom parts of the
source.

\begin{figure}[htbp]
\vspace{8.0cm} \includegraphics{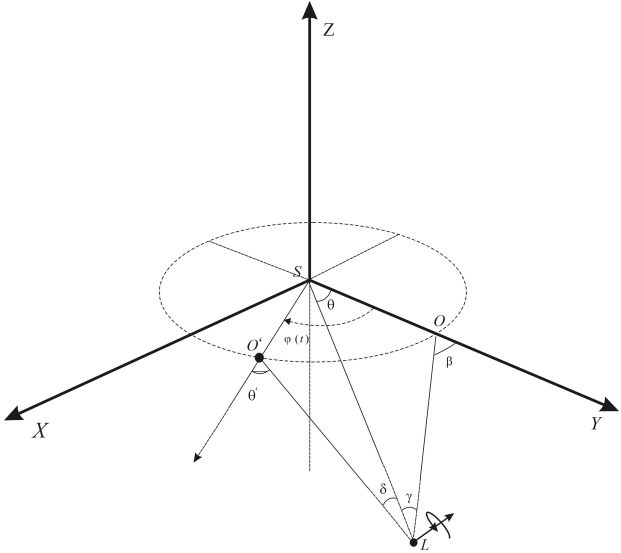} \caption{The geometry of a
retro-lensing event is shown in the case of a general
configuration for the lens, source and observer positions. Photons
emitted by the Sun $S$, in the center of the reference frame, move
towards the retro-MACHO $L$ (a Schwarzschild or a Kerr black hole)
which, for simplicity, is placed on the $Z-Y$ plane. Light rays,
which interact with the black hole strong gravitational field, may
suffer a path deflection to bending angle multiples of $\pi$ and
reach the observer $O^{'}$. The arrow on the lens represents the
black hole spin vector which lies perpendicular to the
lens-observer-source plane.} \label{figure2}
\end{figure}

However, in the general case the source, the observer and the
black hole are not aligned. We consider a reference frame centered
at the source $S$ and with axes $X$ and $Y$ on the Earth orbital
plane (see Fig. \ref{figure2}). We choose $Y$ to contain the lens
in the $Z-Y$ plane and define the misalignment angle $\beta$  as
the angle between the observer-lens direction
($\overrightarrow{OL}$) and the $Y$ axis. Note that the case with
$\beta=0$ and Earth in the orbit position $O$ corresponds to
perfect alignment.

As the Earth moves on its orbit,  the image rings deform into arcs
whose angular extent is $\Delta \Theta \simeq 2 \tan^{-1}
(R_S/D_{OS}~\sin \theta \arcmin)$, where $\theta \arcmin$ is the
angle between the source-observer and observer-lens
($\overrightarrow{SO \arcmin}$ and $\overrightarrow{O \arcmin L}$)
directions, respectively.

The total area of each image is thus given by (Holz \& Wheeler
\cite{hw})
\begin{equation}
A_I \simeq\pi (b_o^2-b_i^2)\frac{\Delta
\Theta}{2\pi}=(b_o^2-b_i^2)\tan ^{-1} \left( \frac{R_S}{D_{OS}\sin
\theta \arcmin }\right) \label{ringimage}~,
\end{equation}
where $b_o$ and $b_i$ represent the outer and inner impact
parameters corresponding to bending angles
\begin{equation}
\alpha _{d,B}=(\pi-\delta) \mp \alpha~,~~~~~\alpha
_{d,S}=(\pi+\delta) \mp \alpha \label{Bangoli}
\end{equation}
for the bigger and smaller image, respectively. Here $\delta$ is
the angle between the $\overrightarrow{SL}$ and
$\overrightarrow{O\arcmin L}$ directions \footnote{Of course,
since the observer moves around the Sun on a circular orbit of
radius $D_{O\arcmin S}\simeq 1~AU$ with angular velocity $\omega
_E\simeq 2\pi/T_E$ (being $T_E$ the orbital period), the Earth
position angle $\phi$ and the angles $\delta$ and $\theta \arcmin$
(see Fig. \ref{figure2}) are time dependent.} (i.e. the deflection
angle needed for a source photon to reach the observer).

The amplification of each image is given by the ratio between the
area of the image given by eq. (\ref{ringimage}) to that of the
source $A_S=R_S^2/(4D_{OS}^2)$, i.e.
\begin{equation}
\mu_{B,S}\simeq (b_o^2-b_i^2)\tan ^{-1} \left(
\frac{R_S}{D_{OS}\sin \theta \arcmin }\right)
 \frac{D_{OS}^2} {\pi D_{OL}^2R_S^2}~.
\label{amplimisali}
\end{equation}
Finally, setting $b_o$ and $b_i$ to the appropriate impact
parameters, the total image amplification is
\begin{equation}
\mu =\mu_B+\mu_S~. \label{totalamplimisali}
\end{equation}
Obviously, in the evaluation of the total amplification $\mu$ we
need the knowledge of the angles $\theta \arcmin$ and $\delta$
defined above which, from inspection of Fig. \ref{figure2}, are
given by
\begin{eqnarray}
\theta \arcmin (t)&=& \arccos \left( \frac{
\overrightarrow{SO\arcmin} \cdot \overrightarrow{O\arcmin
L}}{D_{O\arcmin S}~D_{O\arcmin L}} \right)~,\nonumber \\
\delta (t) &=& \arcsin \left( \frac{D_{O\arcmin S}}{D_{O\arcmin
L}} \sin (\beta-\gamma) \right)~, \label{angoli}
\end{eqnarray}
where $\gamma=\arcsin(D_{OS}\sin \beta /D_{SL})$.

At this stage, we only need to evaluate the photon impact
parameter values $b_o$ and $b_i$ so that light rays coming from
the source are deflected toward the observer.
For a Schwarzschild black hole with mass $M$, Chandrasekhar
(\cite{chandra}) has found that, in the case of large deflection
angle $\alpha_d$, the photon impact parameter, in geometrical
units $G=c=1$, is given by
\begin{equation}
b(\alpha_d)=
b_{1,S}+b_{2,S}e^{-\alpha_d}~,\label{impactparametersch}
\end{equation}
where \begin{equation}
\begin{array}{l}
b_{1,S}=3\sqrt{3}M\\
b_{2,S}=648\sqrt{3}e^{-\pi}M(\sqrt{3}-1)^2/(\sqrt{3}+1)^2.
\label{chandra}
\end{array}
\end{equation}
Consequently, for each position of the Earth around the Sun it is
possible to evaluate the angles $\theta \arcmin (t)$ and $\delta
(t)$, the bending angles $\alpha _{d,B}$ and $\alpha _{d,S}$  and,
through eqs. (\ref{amplimisali}) and (\ref{impactparametersch}),
the total image amplification (\ref{totalamplimisali}) as seen by
the observer.

\begin{figure}[htbp]
\vspace{6.0cm} \includegraphics{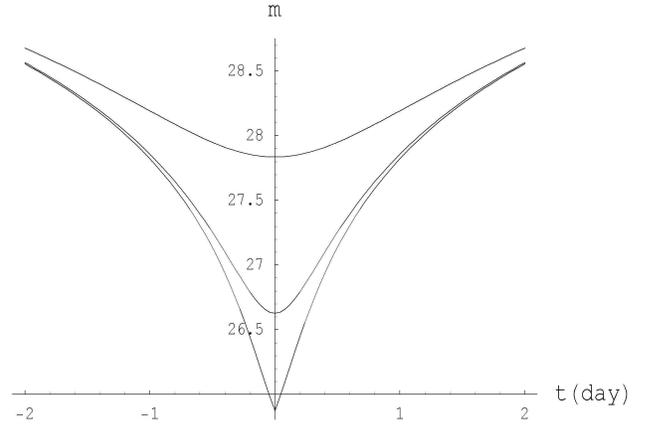} \caption{The expected visual magnitude $m$
of the arcs forming close to a $10~M_{\odot}$ Schwarzschild black
hole (with $a=0$ at $0.01$ pc from the Earth) as a function of
time (i.e of the Earth position) is shown. The curves correspond
to different displacement angles $\beta$. In particular, from top
to bottom, we set $\beta=1\degr$, $\beta=R_{\odot}/1~AU$ and
$\beta=0\degr$, respectively. Note that $t=0$ means, referring to
Fig.\ref{figure2} , that the Earth is on the $Y$ axis.}
\label{figure7}
\end{figure}

The expected light curves (i.e. magnitudes $m$ corresponding to a
certain amplification $\mu$ as a function of time) for a
retro-lensing event due to a Schwarzschild black hole are shown in
Fig. \ref{figure7} for different misalignment angles $\beta$.

\section{Kerr retro-MACHOs}

The results by Holz \& Wheeler (\cite{hw}) on retro-lensing events
hold in the case of Schwarzschild black holes. However, in
general, a black hole is characterized by a non-zero intrinsic
angular momentum which breaks the spherical symmetry and affects
the gravitational field around the compact object.

Indeed, black holes are the ultimate stage of stellar evolution
and may form by SNII explosions. Since stars rotate, angular
momentum conservation implies that black holes also rotate. Black
hole rotation is also confirmed by recent $X$-ray observations of
Fe line width (Zakharov et al. \cite{zaklinee}). Moreover, super
massive black holes at the center of QSOs, AGNs and galaxies show
beamed jet emission implying that they have non zero angular
momentum.

Thus, due to black hole angular momentum $J$, we expect a
modification in the phenomenology of both the usual gravitational
lensing and the retro-MACHO microlensing. Such modifications are
evaluated by a second order expansion of the photon deflection
angles in the spin parameter $a=J/(McR_{\rm Sch})$. Here, $R_{\rm
Sch}=2GM/c^2$ is the black hole Schwarzschild radius. \footnote{
Note that the definition we use for the rotation parameter $a$
corresponds to half of the value in the usual notation
$a=2J/(McR_{\rm Sch})$. Indeed, following Bozza (\cite{bozza2}) we
measure distances in units of the Schwarzschild radius and not in
units of the gravitational radius as is usually done (see e.g.
Shapiro \& Teukolsky \cite{st}).} We consider light trajectories
on the black hole equatorial plane and a more general treatment
will be addressed elsewhere (Zakharov et al. \cite{trajectories}).
However, we expect that the maximal effect due to the black hole
spin occurs for photons lying in the equatorial plane while for
partially inclined orientations the effect becomes closer to the
Schwarzschild case (that is recovered for $\theta=0$).

A simple and reliable method to investigate the subject has been
recently proposed by Bozza (\cite{bozza2}) (for an exhaustive
review of the topics see also Bozza \cite{bozza1}) who revisited
the Schwarzschild and Kerr black hole lensing in a strong
gravitational field.

Considering light rays on the black hole equatorial plane
($\theta=\pi/2$), the Kerr line element in Boyer-Lindquist
coordinates ($t$, $x$, $\theta$, $\phi$) is given by
\begin{equation}
ds^2=A(x)dt^2-B(x)dx^2-C(x)d\phi ^2 +D(x)dtd\phi~,
\end{equation}
with
\begin{equation}
\begin{array}{lll}
\displaystyle{A(x)=1-\frac{1}{x}}\\
\displaystyle{B(x)= \left( 1- \frac{1} {x} + \frac{a^2} {x^2} \right) ^ {-1}} \\
\displaystyle{C(x)=x^2+a^2+\frac{a^2}{x}}\\
\displaystyle{D(x)=2\frac{a}{x}}~,
\end{array}
\end{equation}
where $x$ is the distance from the spinning black hole measured in
units of the Schwarzschild radius in natural units $R_{\rm
Sch}=2M$ and $a=J/(2M^2)$.

Following Bozza (\cite{bozza2}) the impact parameter $b$ of light
rays approaching the black hole is
\begin{equation}
b=\frac{-D_0+\sqrt{4A_0C_0+D_0^2}}{2A_0}~. \label{parimpactx0}
\end{equation}
Here and in the following all the metric functions with the
subscript $0$ are evaluated at $x=x_0$ which is the photon minimum
distance from the black hole.

After a straightforward calculation, the whole deflection angle
(i.e. the bending angle for a photon with impact parameter $b$) is
given by
\begin{equation}
\alpha_d=2\int_{x_0}^{\infty}\frac{d\Phi}{dx}~dx~-\pi,
\label{integrale}
\end{equation}
where
\begin{equation}
\displaystyle{\frac{d\Phi}{dx}=P_1(x,x_0)P_2(x,x_0)}~,
\end{equation}
with
\begin{equation}
\begin{array}{l}
\displaystyle{P_1(x,x_0)=\frac{\sqrt{B}(2A_0Ab+A_0D)}{\sqrt{CA_0}\sqrt{4AC+D^2}}}\\
\displaystyle{P_2(x,x_0)=\left(A_0 -A\frac{C_0}{C}
+\frac{b}{C}(AD_0 - A_0D)\right)^{-1/2}}~. \label{p1p2}
\end{array}
\end{equation}

The deflection angle $\alpha_d$ grows as $x_0$ decreases diverging
at the radius of the photosphere $x_m$ (to be defined below). In
this case the photon is captured by the black hole. Defining the
new variable $z=(x-x_0)/x$  and hence $x=x_0/(1-z)$, the integral
in eq. (\ref{integrale}) becomes
\begin{equation}
\alpha_d=\int_{0}^{1}R(z,x_0)f(z,x_0)~dz~-\pi, \label{integrale2}
\end{equation}
with
\begin{eqnarray}
R(z,x_0)&=&2\frac{(1-y_0)}{A\arcmin (x)}P_1(x,x_0)\nonumber\\
f(z,x_0)&=&P_2(x,x_0)~.
\end{eqnarray}

\begin{figure}[htbp]
\vspace{6.0cm} \includegraphics{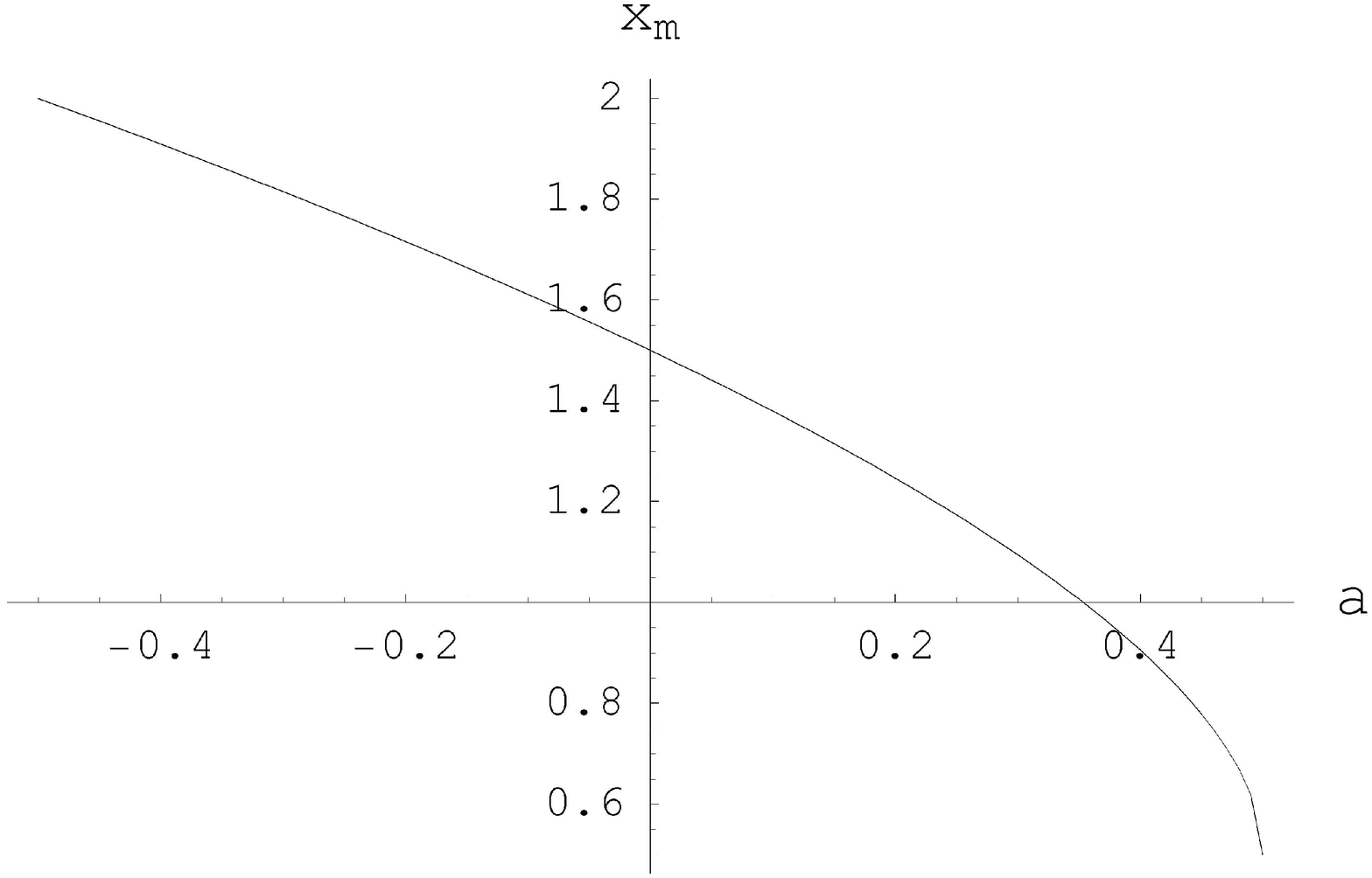} \caption{The radius of the
photosphere as a function of the black hole spin parameter $a$ is
shown. As expected, for positive values of $a$ (i.e. for light
rays co-rotating with the black hole) photons are allowed to get
closer to the black hole. For $a\rightarrow 0$ the photosphere
radius approaches the last stable orbit radius value for a
Schwarzschild black hole ($x_m=3/2$).}
            \label{figure3}
   \end{figure}

The function $R(z,x_0)$ is regular for any values of $z$ while the
function $f(z, x_0)$ diverges as $z\rightarrow 0$. To find the
divergence order, the argument of the square root in $f(z, x_0)$
may be expanded in powers of $z$ obtaining (Bozza \cite{bozza2})
\begin{equation}
f_0(z, x_0)=\frac{1}{\sqrt{c_1(x_0,a) z+c_2(x_0,a)z^2}}~,
\label{f0}
\end{equation}
where $c_1(x_0,a)$ and $c_2(x_0,a)$ represent the first and second
order coefficients of the expansion. Therefore, when $c_1(x_0,a)$
is different from zero the leading order term in the integrand is
$\propto z^{-1/2}$, so that the integral in eq. (\ref{integrale2})
converges. Instead, the integral diverges in the case
$c_1(x_0,a)=0$, corresponding to photons being captured by the
black hole. Accordingly, the photosphere radius $x_m$ is defined
as the outermost solution of equation $c_1(x_0,a)=0$. This
condition is equivalent to $8a^2-x_0(3-2x_0)^2=0$ (for details see
Bozza \cite{bozza2}). Of the three solutions of the cubic equation
above, one is irrelevant as it lies inside the horizon while the
other two correspond to the co-rotating and counter-rotating
photons. The counter-rotating case may also be obtained by taking
$a<0$. The roots are displayed in Fig. \ref{figure3}.

Once the photosphere radius $x_m$ (for a given value of $a$) is
known, one can evaluate the deflection angle $\alpha _d$ from eq.
(\ref{integrale2}) . Thus, retaining only terms up to
$O(x_0-x_m)$, one obtains
\begin{equation}
\alpha_d(x_0)=-k_1 \ln
\left(\frac{x_0}{x_m}-1\right)+k_2+O(x_0-x_m)~, \label{alphax0}
\end{equation}
where
\begin{equation}
\begin{array}{ll}
\displaystyle{k_1=\frac{R(0,x_m)}{\sqrt{c_2(x_m,a)}}}\\
\displaystyle{k_2=-\pi +k_1\ln \frac{2(1-y_m)}{A\arcmin (x_m)
x_m}}+I_R\\
I_R = \displaystyle{\int_0^1
[R(z,x_m)f(z,x_m)-R(0,x_m)f_0(z,x_m)]~dz}~.
\label{k1k2ir}
\end{array}
\end{equation}
The impact parameter $b$ in eq. (\ref{parimpactx0}) can now be
expanded in powers of $(x_0-x_m)$ up to the second order term (the
first order coefficient vanishes) obtaining
\begin{equation}
b \simeq b_{0}(x_m)+b_{2}(x_m)(x_0-x_m)^2\label{parimpactx0sv}~.
\end{equation}

Eliminating the closest approach distance $x_0$ in eqs.
(\ref{alphax0}) and (\ref{parimpactx0sv}) and rearranging the
terms one finally gets (Bozza \cite{bozza2})
\begin{equation}
\alpha_d(b)=-\overline{k}_1 \ln
\left(\frac{b}{b_0(x_m)}-1\right)+\overline{k}_2 +O(b-b_0(x_m))~,
\label{alphadKerrb}
\end{equation}
with
\begin{equation}
\begin{array}{l}
\displaystyle{\overline{k}_1=\frac{k_1}{2}}~,\\
\displaystyle{\overline{k}_2=k_2+\frac{k_1}{2}\ln
\frac{b_2(x_m)x_m^2}{b_0(x_m)}}~.
\end{array}
\end{equation}

\begin{figure}[htbp]
\vspace{6.0cm} $\begin{array}{c} \includegraphics{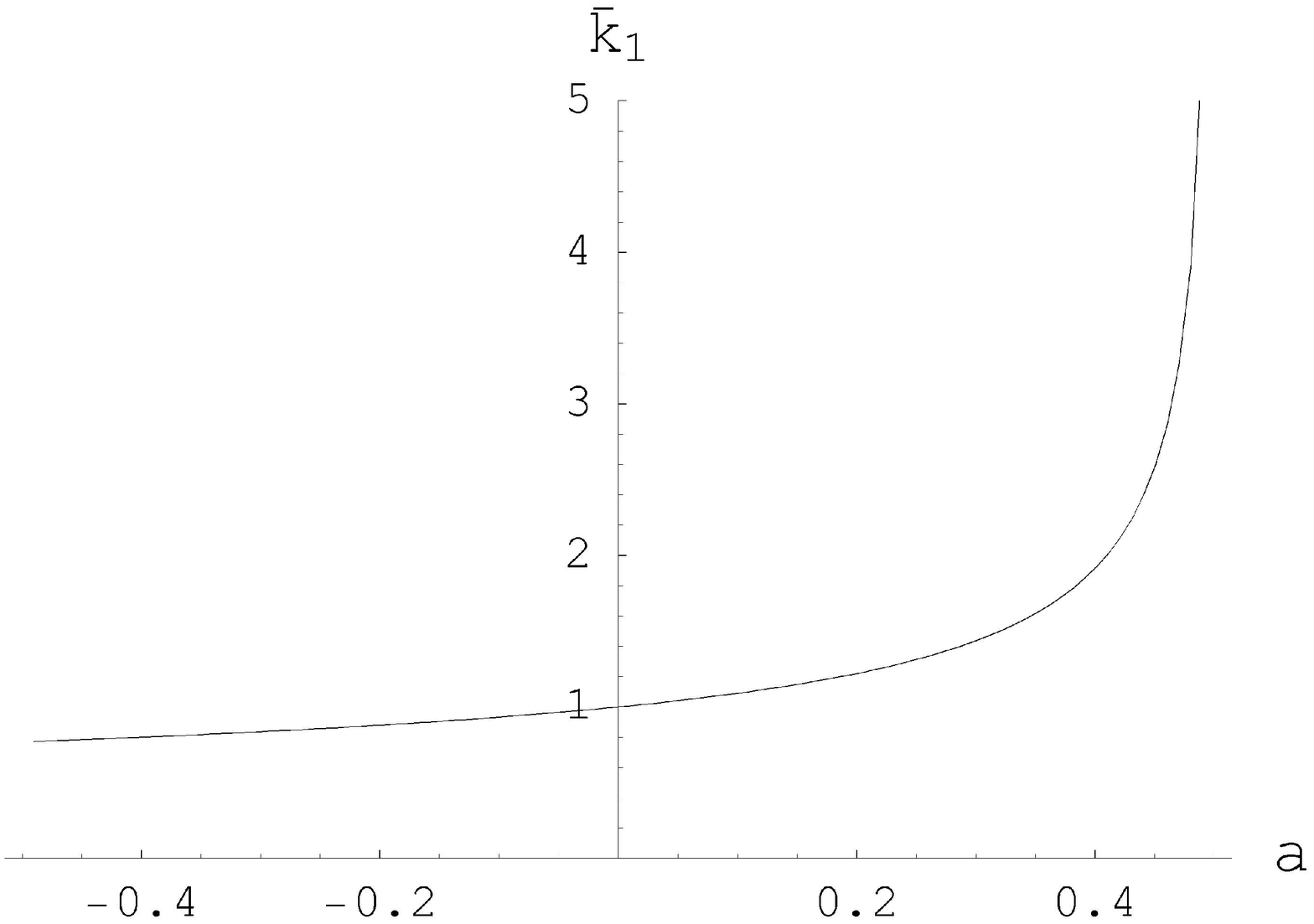}\\[6.0cm]
\includegraphics{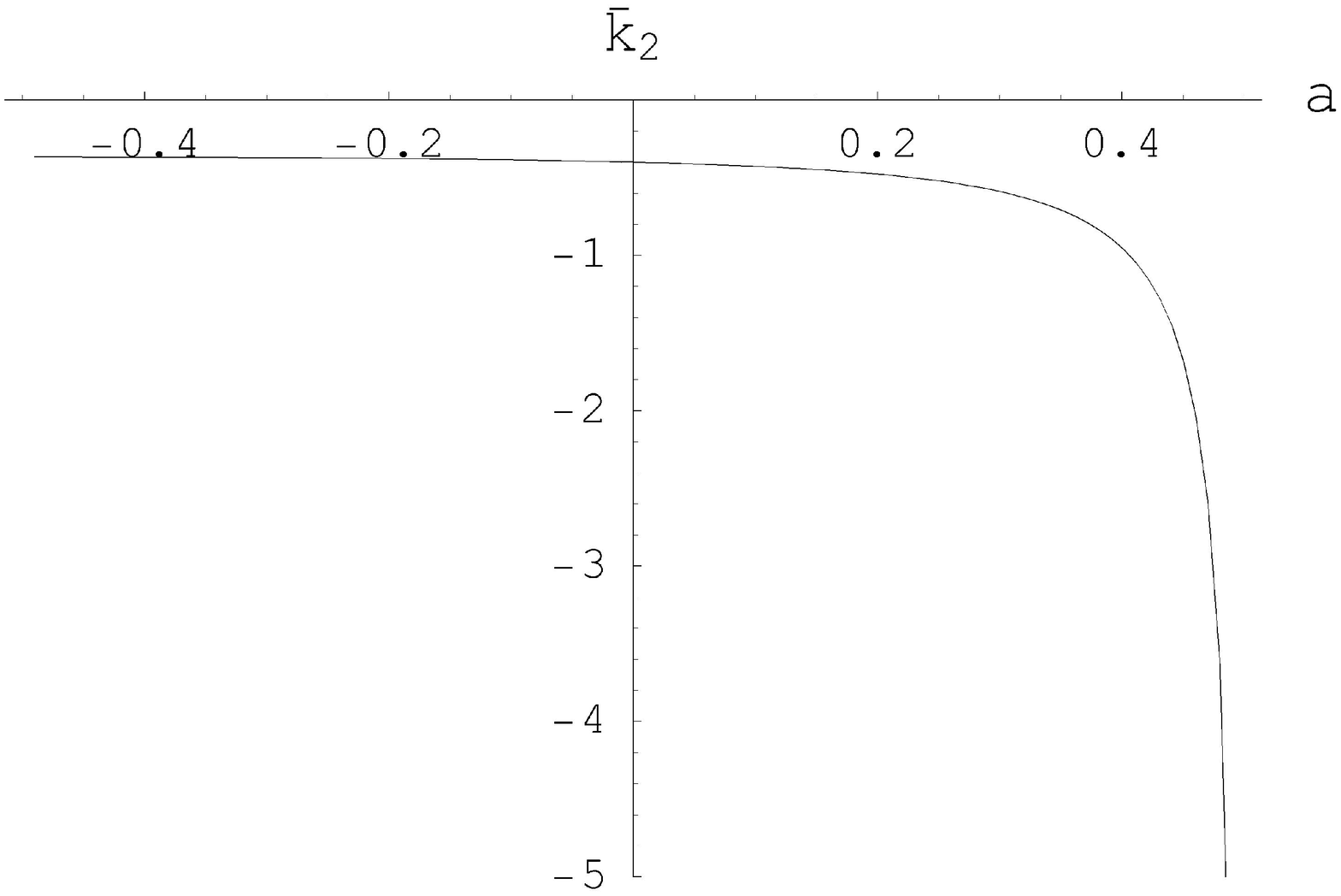}
\end{array}$
\caption{The coefficients $\overline{k}_1$ (upper panel)  and
$\overline{k}_2$ (bottom panel) appearing in eq.
(\ref{alphadKerrb}) are plotted as functions of the black hole
spin parameter $a$. For details see text.} \label{figure5}
\end{figure}

In Fig. \ref{figure5} we  plot the coefficients $\overline{k}_1$
and $\overline{k}_2$ as a function of the black hole spin
parameter. We note that these coefficients diverge for
$a\rightarrow 0.5$ implying that eq. (\ref{alphadKerrb}) can be
used only for low enough values of $a$ for which $x_0 -x_m\ll 1$
holds.

Therefore, by solving eq. (\ref{alphadKerrb}) with respect to the
impact parameter $b$, one finally obtains the generalization
 of the Chandrasekhar relation for Schwarzschild black holes given
in eq. (\ref{impactparametersch}) to Kerr black holes
\begin{equation}
b(\alpha_d)\simeq b_{1,K}+b_{2,K}e^{- \alpha_d/\overline{k}_1}~,
\label{impactparameterkerr}
\end{equation}
with
\begin{equation}
b_{1,K}=2b_0(x_m) M~,~~~~~~~ b_{2,K}=2b_0(x_m) M
e^{-\overline{k}_2/\overline{k}_1}~. \label{aaa}
\end{equation}

Note that the factor $2$ in the previous relations (\ref{aaa}) has
been introduced to make our expression comparable with that in eq.
(\ref{impactparametersch}). It is also easy to verify that for
$a\rightarrow 0$, eq. (\ref{impactparameterkerr}) reduces to the
result (\ref{impactparametersch}) given by Chandrasekhar
(\cite{chandra}) for Schwarzschild black holes.

\begin{figure}[htbp]
\vspace{6.0cm} \includegraphics{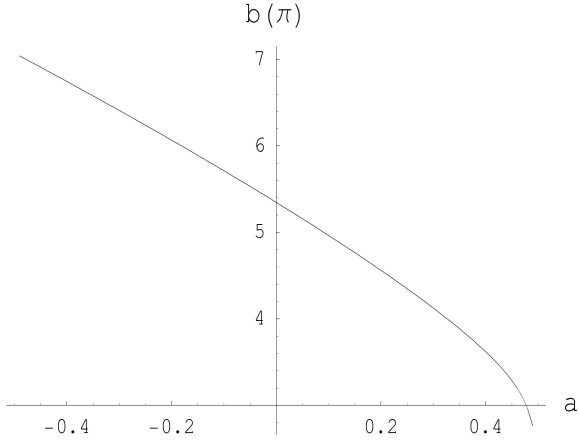} \caption{The impact parameter
$b$, evaluated for the deflection angle $\alpha_d=\pi$, is shown
as a function of the black hole spin parameter $a$. Note that for
$a\rightarrow 0$ the impact parameter approaches the result
$b=5.34664$ for a Schwarzschild black hole given by Chandrasekhar
(\cite{chandra}).}
            \label{figure6}
   \end{figure}

We are now ready to apply this formalism to the retro-lensing
phenomenon. In Fig. \ref{figure6}, assuming $\alpha _d =\pi$, we
plot the impact parameter $b(\pi)$ as a function of the black hole
spin parameter $a$. As one can see, for co-rotating photons
($a>0$) the impact parameters decreases with respect to the
Schwarzschild case (i.e. photons may get closer to the black
hole).

Let us now consider the Sun as the light source and a black hole
of mass $10~M_{\odot}$ at a distance of $0.01$ pc. Since the solar
luminosity in the optical band is $L_{\odot} \simeq 4 \times
10^{33}$ erg $s^{-1}$, corresponding to an optical visual
magnitude $m_{\odot}=-26.8$, the expected magnitude $m$ of the
retro-lensed image can be evaluated as
\begin{equation}
m= m_{\odot}-2.5\log \mu~,
\end{equation}
$\mu$ being the total amplification given by eq.
(\ref{totalamplimisali}). In the case of a Schwarzschild black
hole with $a=0$,  the expected apparent magnitude is given in Fig.
 \ref{figure7} as a function of the time $t$ (corresponding to the
Earth position on its orbit), for different values of the
misalignment angle $\beta$ (see also Holz \& Wheleer \cite{hw}).

\begin{figure*}

\vspace{11.0cm} \includegraphics{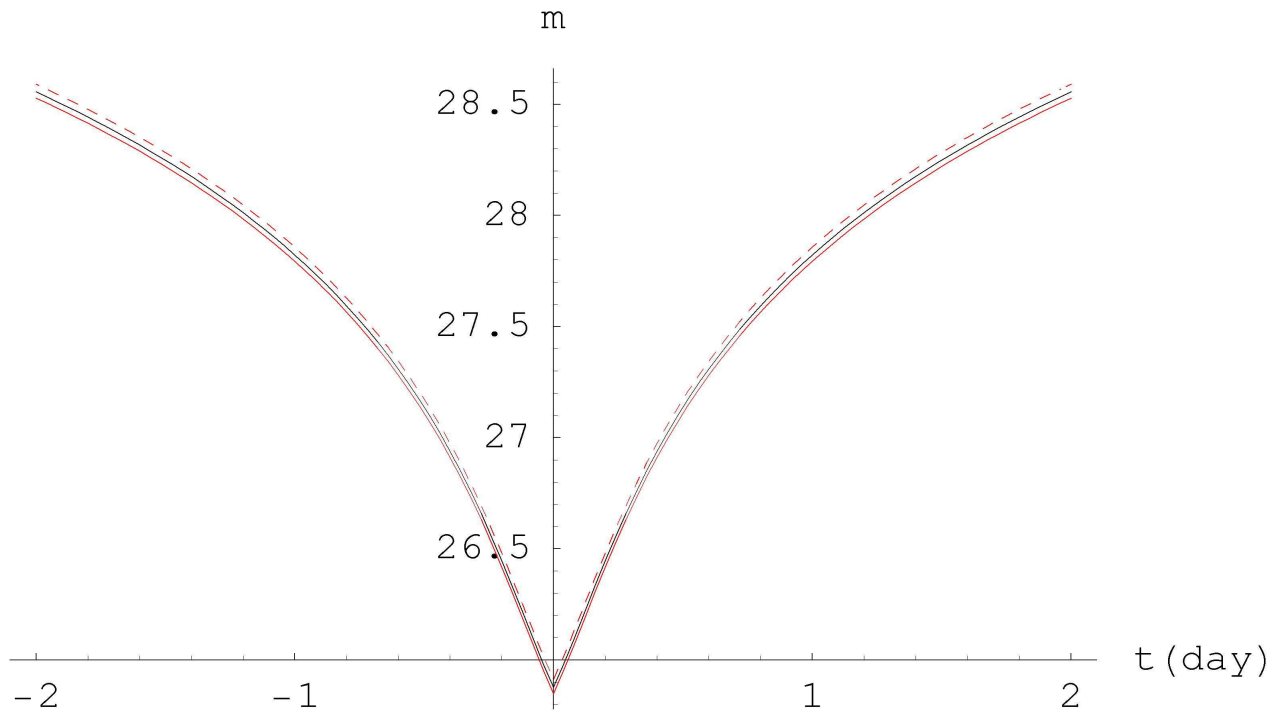} \caption{The expected
retro-lensing light curves for a Kerr black hole with mass
$10~M_{\odot}$ and spin parameter $a=0.1$ is given as a function
of the time (i.e. of the Earth position in its orbit). The solid
red line corresponds to the magnitude of the image formed by
co-rotating photons while the dashed red line is the magnitude of
the image formed by counter-rotating photons. The central black
line corresponds to a retro-lensing event involving a
Schwarzschild black hole with the same mass. The light source is
the Sun and the black hole is placed at $0.01$ pc from Earth.}
\label{figure8}

\end{figure*}

In the case of a Kerr black hole we can evaluate the Sun image
amplification by using eqs. (\ref{amplimisali}) -(\ref{angoli}),
in the most favorable case of perfect alignment ($\beta = 0$). For
this purpose we only need to evaluate, through eq.
(\ref{impactparameterkerr}), the impact parameters corresponding
to the bending angles for which light rays emitted by the Sun may
reach the observer after deflection of the angles $\alpha_{d,B}$
and $\alpha_{d,S}$ defined in Section 2. For a $10~M_{\odot}$ Kerr
black hole with spin parameter $a=0.1$ at a distance of $0.01$ pc
and misalignment angle $\beta =0$, the expected retro-lensing
light curve is given in Fig. \ref{figure8} (red solid line). In
the same Figure we also give the light curve (dashed red line)
corresponding to counter-rotating photons. As expected, for a Kerr
black hole with spin parameter $a$, only a modification up to the
second order in the retro-MACHO amplification appears, as is
evident by inspecting Fig. \ref{figure8} where the lightcurves for
the Kerr black hole case are reported (dashed and solid red lines)
together with the corresponding one for a Schwarzschild black hole
with the same mass and $\beta = 0$ (central black solid line).
\footnote{Co-rotating and counter-rotating photons should produce
an image scintillation due to their relative time delay. However,
the time scale for such scintillations goes from $\sim 10^{-5}$ s
to $\sim 10$ s for black holes with mass in the range $1$
M$_{\odot}$-$10^6$ M$_{\odot}$. Due to the short scintillation
time scale with respect to the integration time required by
observation, the net result is simply that of seeing the average,
which coincides with the light curve for the Schwarzschild black
hole.}  We also notice that we have neglected multiple rotations
of the photons around the black hole. As emphasized by Holz \&
Wheeler (\cite{hw}), photons may form a series of deformed rings
corresponding to deflection angles that differ by $2n\pi$ (with
$n$ as integer number). However, it is clear that the brightest
image is that corresponding to the outermost ring (obtained for
$\alpha_d\simeq\pi$) so that neglecting the other images does not
change substantially the retro-lensing image brightness.

\section{Discussion}

It has been shown by Holz \& Wheeler (\cite{hw}) that the limiting
distance $D_L$ at which the rings around a black hole of mass $M$
illuminated by our Sun can be observed is given by
\begin{equation}
D_{L} =0.02 {\rm pc} \times \left[10^{(m-30)/2.5}(M/10~
M_{\odot})^2\right]^{1/3}~, \label{con_1}
\end{equation}
where the ring magnitude of the retro-lensing event has been
chosen to be $30$ (see Fig. \ref{figure7}). For example, if the
Sun is the light source and the black hole has mass $10$
M$_{\odot}$, the limiting distance at which the retro MACHO can be
detected is of only $0.02$ pc for an instrument with $m=30$ (such
as HST). On the other hand, it is very unlikely that black holes
so close to the Earth exist. However, the most convenient geometry
to get luminous rings around a black hole would be that of a very
bright star close to a black hole at some distance $D_L$ from
Earth. \footnote{It has been shown that the best chance of
observing retro-lensing images in the future is by looking towards
the galactic center black hole in Sgr A$^*$ around which a very
bright star (named S2) with a mass of about $15~M_{\odot}$ is
orbiting. The resulting magnitude of the retro-lensing images in
the Schwarzschild case are in the range $33.3-36.8$ (depending on
the star distance from the black hole) in the K-band, close to the
limiting magnitude of the next generation of space-based
telescopes (De Paolis et al. \cite{dgin}.)} In this case, and for
the most general geometrical configuration of source, lens and
observer with relative distances $D_{OL}$, $D_{LS}$ and $D_{OS}$,
the instrumental limiting magnitude $\bar{m}$ necessary to see the
rings is thus given by
\begin{equation}
\bar{m}\geq m_{\lambda} -2.5\log \mu ~, \label{generale}
\end{equation}
$m_{\lambda}$ being the magnitude of the source in a certain
electromagnetic band and $\mu$ the ring amplification given by eq.
(\ref{totalamplimisali}). In the particular case of perfect
alignment between observer, lens and source, the previous relation
reads
\begin{equation}
\bar{m}\geq m_{\lambda} -2.5\log
\left[1.04\left(\frac{M}{M_{\odot}}\right)^2\left(\frac{R_{\odot}}{R_S}\right)
\frac{D_{OS}^2}{D_{LS}D_{OL}^2}\right]~,
\end{equation}
which reduces to eq. (\ref{con_1}) taking the geometry adopted by
Holz \& Wheeler (\cite{hw}). Notice that in the previous equation
we are considering all the distances ($D_{OS},D_{LS}$ and
$D_{OL}$) measured in cm.

As is clear from Fig. \ref{figure8}, the spin effect on the image
brightness is a minor effect and so it cannot increase
substantially the image amplification with respect to the
Schwarzschild black hole case.  Moreover, as already stated, our
analysis is confined to photons lying in the black hole equatorial
plane and the retro-lensing image magnitude is calculated by using
eq. (\ref{totalamplimisali}). This equation is rigorously valid
only in the Schwarzschild case since it assumes that each part of
the image is isoluminous (and this is not true for Kerr black
holes). However, if we limit the estimate to only the total
magnitude of the retro-lensing images, we do not expect a
substantial change with respect to the Schwarzschild case since
the increase in the luminosity of the corotating edge should
approximately compensate the luminosity decrease of the
counter-rotating edge.

We emphasize again that our treatment is valid only for slowly
rotating black holes but it is clear that increasing the rotation
parameter $a$, the brightness of the retro-lensing images also
increases. A complete (purely numerical) treatment of this issue -
valid for arbitrarily fast rotating black holes - will be
published elsewhere (De Paolis et al. \cite{kerrcomplete}).

Since in the slow rotation limit the light curves are almost
indistinguishable from the Schwarzschild case, the most important
effect of the black hole spin is that of deforming the ring shape
with respect to the Schwarzschild case (Peter \cite{peter76}, Bray
\cite{bray}). A detailed analysis of this effect will be presented
elsewhere (Zakharov et al. \cite{trajectories}). Detecting the
shape of the ring images might give a unique possibility of
investigating the black hole parameters (spin $a$, mass $M$ and
relative position of observer-lens-black hole) in an active way.

To really observe the ring shape, the detecting instrument must
have a high enough angular resolution in the range $1-10$
$\mu$arcsec for most cases of interest. Future space-based
instruments like the Space Interferometry Mission (SIM) (see Shao
\cite{sim} for technical details), GAIA Satellite (see e.g.
Belokurov \& Evans \cite{gaia}), FAME (Johnston et al.
\cite{fame}) and MAXIM (\cite{maxim}) would have the angular
resolution capabilities to resolve such images and therefore will
be able to observe retro-MACHOs if the instrumental limiting
magnitude satisfies eq. (\ref{generale}).

\end{document}